\documentclass[reprint,amsmath,amssymb,aps,showpacs]{revtex4-1}
\usepackage{graphicx}% Include figure files
\usepackage{dcolumn}% Align table columns on decimal point
\usepackage{bm}% bold math
\usepackage[colorlinks,linkcolor=blue,citecolor=blue]{hyperref} 
\usepackage{epstopdf}
\def\be{\begin{equation}}

\def\ee{\end{equation}}
\def\barr{\begin{array}}
\def\earr{\end{array}}

\def\l{\left}
\def\r{\right}
\def\dis{\displaystyle}
\def\ed{\end{document}}

\begin{document}

\preprint{APS/123-QED}

\title{Generalized Gaussian wave packet dynamics: Integrable and Chaotic Systems}

\author{Harinder Pal}
\affiliation{Department of Physics and Astronomy, Washington State University, Pullman, WA, USA 99164-2814}
\affiliation{Instituto de Ciencias F\'{i}sicas, Universidad Nacional Aut\'{o}noma de M\'{e}xico - C.P. 62210 Cuernavaca, M\'{e}xico}

\author{Manan Vyas}
\affiliation{Department of Physics and Astronomy, Washington State University, Pullman, WA, USA 99164-2814}
\affiliation{Instituto de Ciencias F\'{i}sicas, Universidad Nacional Aut\'{o}noma de M\'{e}xico - C.P. 62210 Cuernavaca, M\'{e}xico}

\author{Steven Tomsovic}
\affiliation{Department of Physics and Astronomy, Washington State University, Pullman, WA, USA 99164-2814}
\affiliation{Max-Planck-Institut f\"ur Physik komplexer Systeme, N\"othnitzer Stra$\beta$e 38, D-01187 Dresden, GE}

\date{\today}

\begin{abstract}

The ultimate semiclassical wave packet propagation technique is a complex, time-dependent WBK method known as generalized Gaussian wave packet dynamics (GGWPD).  It requires overcoming many technical difficulties in order to be carried out fully in practice.  In its place roughly twenty years ago, linearized wave packet dynamics was generalized to methods that include sets of off-center, real trajectories for both classically integrable and chaotic dynamical systems that completely capture the dynamical transport.  The connections between those methods and GGWPD are developed in a way that enables a far more practical implementation of GGWPD.  The generally complex saddle point trajectories at its foundation are found using a multi-dimensional, Newton-Raphson root search method that begins with the set of off-center, real trajectories.   This is possible because there is a one-to-one correspondence. The neighboring trajectories associated with each off-center, real trajectory form a path that crosses a unique saddle; there are exceptions which are straightforward to identify.  The method is applied to the kicked rotor to demonstrate the accuracy improvement as a function of $\hbar$ that comes with using the saddle point trajectories. 

\end{abstract}

\pacs{03.65.Sq,05.45.-a,05.45.Mt}

\maketitle

\section{Introduction}
\label{Intro}

Gaussian wave packet propagation is an extremely important tool for understanding a vast array of physical problems.  
For example, it has been applied to problems involving driven cold atoms~\cite{Bakman15}, electrons in strong fields~\cite{Zagoya14}, fidelity studies~\cite{Jalabert01,Cerruti02}, and a broad range of spectroscopic and pump-probe experiments~\cite{Heller81b,Alber91}.  In the short wavelength limit, it is natural to rely on semiclassical methods and approximations as a way of understanding the essential physics and for making practical calculations.  The pinnacle of such approaches, excluding the introduction of higher order uniformizations of caustics or diffraction corrections, is a complex, time-dependent WBK method.  For Gaussian wave packet propagation that was shown to be  generalized Gaussian wave packet dynamics (GGWPD)~\cite{Huber87,Huber88}.  Although it is not identical, there is substantial overlap between GGWPD and the earlier work on the saddle point approximation applied to the path integral in a coherent state representation~\cite{Klauder78,Weissman83}.  

In GGWPD, classical Lagrangian manifolds of complex phase space points are identified for both initial and final wave packet states; one can also identify the final state as a position or momentum eigenvector and construct the fully propagated wave packets as well.  There is a boundary value problem of identifying the initial complex manifold  of phase points as initial conditions, propagating them, and intersecting the manifold of ending phase points with the final complex manifold.  That determines the generally complex saddle points of an analytically continued, time-dependent WBK theory~\cite{OzorioBook} for propagating wave packets, and that information can be used to generate the best semiclassical approximation to the dynamics (neglecting uniformization and diffraction extensions).  

Direct, full implementation of GGWPD is rather impractical for non-trivial, two and higher degree-of-freedom dynamical systems. There are a few basic reasons for this.  First, for each degree of freedom, there are four dimensions in the complexified phase space.  The initial and final Lagrangian manifolds have half the dimensions.  In one-degree-of-freedom ($D=1$) dynamical systems, the saddle points lie at intersections of two two-dimensional, unbounded manifolds in a four-dimensional space.  For $D=2$ the manifolds are four-dimensional in an eight dimensional phase space.  Thus, increasing the number of degrees of freedom appears to push the boundary value problem beyond treatment.  Second, complexified classical mechanics has many challenging complications related to the analytical structure of the complexified phase space, such as ``runaway'' trajectories that have infinite values of position or momentum within finite times, which are indicative of singularities and branch cuts~\cite{Huber88}.  The determination of whether the contribution of a particular saddle should be included, i.e. is on the good side of a Stokes surface, can be very challenging.  As a final example,  beginning with $D\ge 2$ dynamical systems, there is a likely possibility of at least some chaotic dynamics.  As the length of the propagation time increases, the number of necessary saddle points has to explode exponentially rapidly.  There is recent work on the implementation of GGWPD~\cite{Zamstein14,Zamstein13}, though for the most part, the above mentioned difficulties remain.   

Considerable work on semiclassical wave packet methods has been directed towards developing practical calculational tools, but with some compromises.  Several initial value representations have been developed~\cite{Heller81, Herman84, Heller91b, Kay94}, that effectively are sophisticated, hybrid semiclassical-numerical integration methods.  They have the advantages of simpler implementation than GGWPD and the uniformization of some kinds of caustic singularities.  Several difficulties exist with these methods as well~\cite{Baranger01}, not least of which are that increasing propagation times and numbers of degrees-of-freedom tend to rapidly expand the quantity of necessary trajectories, and some of the methods weight most the least contributing trajectories (that end up largely relying on phase cancellations to reduce their importance).

Another approach is to generalize linear wave packet dynamics~\cite{Heller75,Heller91} to the use of representative, off-center, but real trajectories.  Each representative trajectory, through the quadratic expansion of the classical action function and subsequent integration is incorporating the effects of a subset of an infinite number of like-behaving trajectories found in continuous branches.  In this way, the representative, off-center, real trajectories have the advantage of capturing the complete classical transport, i.e.~all classically allowed dynamical possibilities are identified and incorporated.  Thus, any dynamical nonlinearities are taken into account.  These methods were applied to the hydrogen atom~\cite{Barnes93,Barnes94}, an integrable system, the stadium billiard~\cite{Tomsovic91b,Tomsovic93}, and the quantum bakers map~\cite{Oconnor92}, the latter two systems being fully chaotic.  They proved to be extremely accurate to time scales far beyond the Ehrenfest time~\cite{Oconnor92,Sepulveda92,Tomsovic93} where linearized wave packet dynamics break down.

For integrable systems, the desired off-center, real trajectories are selected from a manifold of real phase points that is normal to the energy surface.  In local action-angle coordinates, the angle variables are fixed on the manifold, and the action values are varied over the part of the phase space that has significant weight in the Wigner transforms of the wave packets.  So long as a single set of action-angle variables is necessary within the wave packet's domain of the phase space, the corresponding shearing trajectory sum captures the complete classical transport.  

For chaotic systems, the trajectory sum is over all trajectories heteroclinic to the initial and final phase space centroids of the wave packets.  These are trajectories that in the infinite past approach the infinite past of the initial wave packet's phase space center, and hence lie on its unstable manifold, and in the infinite future approach the infinite future of the final wave packet's center, and hence lie on its stable manifold.  The intersections of the two manifolds identify the heteroclinic trajectories.  This classical transport problem is completely solved by the resulting heteroclinic trajectory sums in the limit that the corresponding classical densities are well localized within convergence zones in the normal coordinate forms~\cite{Moser56,Silva87}.  

Although, well motivated by physical considerations, off-center, real trajectory methods are not on as solid mathematical foundations as GGWPD or more generally, time-dependent WBK methods~\cite{Maslov81}.  Developing the links between the two methods has multiple purposes.  It sheds light on the missing mathematical foundations of off-center, real trajectory methods, provides an interpretation and a more economical and intuitive means of implementing GGWPD that incorporates the full classically allowed transport, and the $\hbar$-dependence can be analyzed and contrasted for the two methods.  

Each true saddle point found using a representative, off-center, real trajectory corresponds to the classical transport pathway represented by the aforementioned branch of trajectories associated with that off-center, real trajectory (i.e.~its infinite subset of like-behaving trajectories).  Within that branch are real trajectories that together form a path that crosses from one side to the other of the saddle.  Of course, that path doesn't pass through the saddle point, which is complex.  These saddles are always on the good side of Stokes surfaces.  The existence of saddle points other than these within GGWPD would correspond to tunneling corrections or those that must be excluded by being on the wrong side of Stokes surfaces.  

The paper is organized as follows, the next section gives the background information needed for GGWPD and off-center, real trajectory sums for integrable and chaotic systems.  This is followed by a description of an immensely easier implementation method for GGWPD beginning with the presumed known off-center, real trajectories, and a derivation of the saddle point expression.  A very simple, analytical example is worked out in detail to illustrate the comparative workings of successively sophisticated semiclassical approximations; i) linearized wave packet dynamics, ii) off-center, real trajectory methods, and iii) GGWPD.  The GGWPD implementation method is applied to the kicked rotor in a near-integrable regime and another that is strongly chaotic for the purpose of showing the improvement of carrying out GGWPD relative to off-center, real trajectory methods as a function of $\hbar$.  Lastly, we point to interesting directions for future work.

\section{Background}
\label{back}

It is convenient to make use of both Dirac and wave function notations.  Using the greek letters, $\alpha, \beta$  as a shorthand to denote the parameters that define particular kets associated with unit normalized Gaussian wave packets,
\begin{equation}
|\alpha\rangle = | \vec p_\alpha, \vec q_\alpha, {\bf b}_\alpha \rangle\ \,
\end{equation}
where the mean position and momentum parameters, $(\vec p_\alpha, \vec q_\alpha)$, can be regarded as position and conjugate momentum variables in a $2D$-dimensional {\it real} phase space, and ${\bf b}_\alpha$ is a symmetric, positive definite $D$-dimensional variance/covariance matrix; this matrix could have complex entries as long as all of its eigenvalues have positive real parts (complex entries would give the flexibility of building in ``chirping'' of the wave functions).  This leads to
\begin{eqnarray}
\label{wavepacket}
\langle \vec x |\alpha\rangle &=& \langle \vec x| \vec p_\alpha, \vec q_\alpha, {\bf b}_\alpha \rangle = \phi_\alpha(\vec x) \nonumber \\
&=& \left(\frac{2^D{\rm Det}[{\bf b}_\alpha]}{\pi^D}\right)^{1/4}\exp\left[ - \left(\vec x - \vec q_\alpha \right) \cdot {\bf b}_\alpha \cdot \left(\vec x - \vec q_\alpha \right)  \right. \nonumber \\ 
&& \left. +\frac{i}{\hbar} \vec p_\alpha \cdot \left(\vec x - \vec q_\alpha \right)\right]
\end{eqnarray}
where implicitly the right vectors are column vectors and the left vectors are row vectors.  The form of the normalization constant is valid for ${\bf b}_\alpha$ real symmetric, otherwise it has to be replaced by $({\bf b}_\alpha+{\bf b}_\alpha^*)/2^D$; the equations given ahead also assume ${\bf b}_\alpha$ real symmetric.  To obtain the normalization and several of the equations further ahead in the text, a useful integral identity is
\begin{equation}
\label{handy}
\left(\frac{\pi^D}{{\rm Det}[{\bf A}]}\right)^{1/2} = \int_{-\infty}^\infty {\rm d}\vec x \exp\left[ - \left(\vec x - \vec q_0 \right) \cdot {\bf A} \cdot \left(\vec x - \vec q_0 \right)  \right] 
\end{equation}
where the matrix $\bf A$ must satisfy the constraints on ${\bf b}_\alpha$ mentioned above.

For mechanical dynamical systems with potentials of constant, linear, or quadratic spatial dependence, even if time-dependent, the evolution is such that an initial Gaussian wave packet remains a Gaussian, albeit possibly moved and distorted by the dynamics.  Letting $t_0=0$, the unitary evolution of a system governed by a Hamiltonian $H$ gives
\begin{equation}
U_H(t)|\alpha\rangle = |\alpha_t \rangle
\end{equation}
where the time-dependence of the parameters can be solved by making use of the classical trajectory with $(\vec p_\alpha, \vec q_\alpha)$ as its initial condition.  For more general potentials, this relation remains approximately true up to the Ehrenfest time scale~\cite{Ehrenfest27} on which the wave packet remains localized well enough to correspond to a classical density, and is the approximation method known as linearized wave packet dynamics~\cite{Heller75,Heller91}.

For physical problems in which knowing the dynamics approximately up to the Ehrenfest time scale suffices, it is an extraordinarily useful method~\cite{Heller81b}.  The great simplification is that there is no need to determine which trajectory or trajectories to use, the initial condition is $(\vec p_\alpha, \vec q_\alpha)$.  Running this trajectory and solving the stability equations can be done in any number of degrees of freedom as long as the Hamiltonian is known or well approximated.  In some cases, such as found in many spectroscopic experiments, the quantity of interest may be a correlation function (or it can be considered a matrix element of the time-dependent Green function in a wave packet representation or its Fourier transform), $\langle \beta | U_H(t)|\alpha\rangle$.  Even if the means of $\phi_\beta(\vec x), \phi_\alpha(\vec x)$ cannot be connected by a real trajectory, the limitations of using linearized wave packet dynamics are roughly the same, i.e.~it works up to the Ehrenfest time, unless $\phi_\beta(\vec x)$ is far in the tail of $\langle \vec x | U_H(t) |\alpha\rangle$, where the approximation would fail earlier, but the overlap would also be quite small.

Generally speaking, beyond the Ehrenfest time scale nonlinearities arise in the dynamics and the possibility of multiple pathways for wave amplitudes to interfere opens up.  Both phenomena signal the breakdown of a linearization of the dynamics and a more sophisticated method becomes necessary that is capable of accounting for all the nonlinearities and wave interference.  As noted in the introduction, the pinnacle of semiclassical approximations for wave packets is GGWPD~\cite{Huber87,Huber88} and it does contain all the information about nonlinear dynamics in the short wavelength regime.

\subsection{GGWPD}
\label{general}

One way to view the key element of a time-dependent WBK theory is the identification of the appropriate Lagrangian manifold of trajectories for a given state.  Approximate quantum propagation follows by the propagation of the manifold as a set of trajectory initial conditions.  If interest lies in the propagated state's overlap with a final state, then the intersections of the propagated manifold with the manifold of the final state gives the needed stationary phase points.  

The key underlying GGWPD is the identification of this manifold.  If one allows for complex positions and conjugate momenta, then there is an ambiguity in the form of a wave packet as described by Eq.~(\ref{wavepacket}).  Leaving aside the global phase and normalization for a moment, any complexified pair of position, momentum values $(\vec {\cal P}, \vec {\cal Q})$, which preserves the  column vector relation,
\begin{equation}
\label{constraints}
2 {\bf b}_\alpha \cdot \vec {\cal Q} + \frac{i}{\hbar} \vec {\cal P} = 
2 {\bf b}_\alpha \cdot \vec q_\alpha + \frac{i}{\hbar} \vec p_\alpha
\end{equation}
leaves the spatial dependence of the Gaussian wave packets identical.  This can be verified by inspection of the resultant quadratic forms of the exponential argument.  In fact, the set of all $\{(\vec {\cal P}_\alpha, \vec {\cal Q}_\alpha)\}$ satisfying the ket equations is the appropriate Lagrangian manifold underlying $|\alpha\rangle$~\cite{Huber88}.  Clearly, if one were to scale ${\bf b}_\alpha$ with $\hbar^{-1}$, then $\hbar$ drops out of the equation entirely.  This scaling fixes the overall shape of a wave packet's phase space density, only the volume is determined by $\hbar$.  Since the locations of the saddle points are independent of $\hbar$ with this scaling, this gives an excellent approach to studying the accuracy of the method as $\hbar \rightarrow 0$.

Although, all the points on the above manifold leave the spatial dependence invariant, expressing Eq.~(\ref{wavepacket}) and its bra vector version in terms of the complex phase space variables does not leave the normalization and phase invariant.  Consider a particular point on the ket manifold as a complex center of the wave packet, say $(\vec {\cal P}_0, \vec {\cal Q}_0)$.  Adjusting the normalization coefficient to
\begin{eqnarray}
\label{norm1}
{\cal N}_\alpha^0 &=& \left(\frac{2^D{\rm Det}[{\bf b}_\alpha]} {\pi^D}\right)^{1/4} \exp \left[ \frac{i}{\hbar}\left( \vec{\cal P}_0 \cdot \vec{\cal Q}_0 - \vec p_\alpha \cdot \vec q_\alpha\right) + \right.  \nonumber \\
&& \left.  \vec {\cal Q}_0 \cdot {\bf b}_\alpha \cdot \vec {\cal Q}_0 - \vec q_\alpha \cdot {\bf b}_\alpha \cdot \vec q_\alpha \right. \Big]
\end{eqnarray}
restores both the normalization and phase.  This is already indicating something useful about the phase point $(\vec {\cal P}_0, \vec {\cal Q}_0)$.  By using Eq.~(\ref{constraints}) and its complex conjugate form, the real variables $(\vec p_\alpha, \vec q_\alpha)$ can be eliminated from Eq.~(\ref{norm1}).  For the purpose of separating the contributions to the normalization and phase explicitly, it is convenient to have a notation for the real and imaginary parts of the phase points.  Denote them $\vec {\cal P}_0= \vec {\cal P}_0^R + i \vec {\cal P}_0^I$ and $\vec {\cal Q}_0= \vec {\cal Q}_0^R + i \vec {\cal Q}_0^I$, where ahead the $0$ refers to time (initial condition or final condition, $t$).  After some algebra, one finds
\begin{eqnarray}
\label{norm2}
{\cal N}_\alpha^0 &=& \left(\frac{2^D{\rm Det}[{\bf b}_\alpha]} {\pi^D}\right)^{1/4}  \exp \left[ F^-_0(\alpha) \right] \nonumber \\
F^-_0(\alpha) &=&  \frac{i}{2\hbar^2} \vec {\cal P}_0^R \cdot {\bf b}^{-1}_\alpha \cdot \vec {\cal P}_0^I - \frac{1}{4\hbar^2} \vec {\cal P}_0^I \cdot {\bf b}^{-1}_\alpha \cdot \vec {\cal P}_0^I   \nonumber \\
&&   -  \vec {\cal Q}_0^I \cdot {\bf b}_\alpha \cdot \vec {\cal Q}_0^I - \frac{1}{\hbar} \vec {\cal P}^R_0 \cdot \vec {\cal Q}^I_0  
\end{eqnarray}
where the $-$ sign in the notation $F^-_0$ is for a ket vector.  The first term adjusts the phase and the last three terms the normalization.    The bra vector version is given by
\begin{eqnarray}
\label{norm3}
{\cal N}_\beta^t &=& \left(\frac{2^D{\rm Det}[{\bf b}_\beta]} {\pi^D}\right)^{1/4}  \exp \left[ F^+_t(\beta) \right] \nonumber \\
F^+_t(\beta) &=&  \frac{i}{2\hbar^2} \vec {\cal P}_t^R \cdot {\bf b}^{-1}_\beta \cdot \vec {\cal P}_t^I - \frac{1}{4\hbar^2} \vec {\cal P}_t^I \cdot {\bf b}^{-1}_\beta \cdot \vec {\cal P}_t^I   \nonumber \\
&&   -  \vec {\cal Q}_t^I \cdot {\bf b}_\beta \cdot \vec {\cal Q}_t^I + \frac{1}{\hbar} \vec {\cal P}^R_t \cdot \vec {\cal Q}^I_t  
\end{eqnarray}
where three minor changes are necessary.  In Eqs.~(\ref{wavepacket},\ref{constraints}-\ref{norm3}), the sign of the momenta are reversed, $\alpha$ is replaced by $\beta$ for the final wavepacket, and the time is replaced by the final $t$ instead of the initial $t=0$. 

In a number of works, a concept of complex, but {\it nearly real} trajectories is used; see~\cite{Vanvoorhis02,Vanvoorhis03} and references therein.  This has an important intuitive appeal, but there does not appear to be any definition of the meaning of nearly real.  The normalization correction terms of Eqs.~(\ref{norm2},\ref{norm3}) may provide a step in that direction.  Although, there is no measure of distance in real phase space, let alone complex phase space, the sum of terms given by imaginary momenta weighted by ${\bf b}^{-1}_\alpha$, imaginary position weighted by ${\bf b}_\alpha$, and the correlation between the real momentum and imaginary position is unit free.  Furthermore, this sum contains a useful wave-packet-shape and $\hbar$-dependence.  Ahead it is seen to show up directly in the expressions for propagating wave packets and correlations functions, Eqs.~(\ref{psit},\ref{corrt}), as an exponential decay factor.  One might then consider the complexity of any complex phase point $(\vec {\cal P}_0, \vec {\cal Q}_0)$ to be greater, the greater the decay factor given in Eq.~(\ref{norm1}).  The measure of complexity of a phase point is generally not the same for $F^-_0(\alpha)$ and $F^+_t(\beta)$.

However, some caution is warranted and we find that considering a saddle point trajectory to be nearly real can be rather misleading.  Instead, the property of whether or not a saddle has a set of real trajectories that form a path across itself is better defined and more fundamental.  We call such saddles {\it real crossing saddles} and the rest {\it non-crossing} ones.  It is worth noting that the definition of a real crossing saddle does not imply that its saddle point is actually nearly real by any kind of intuitive logic.  There generally are real crossing saddles with very strongly Gaussian damped contributions.

For a quantity such as $\langle \beta | U_H(t)|\alpha\rangle$, the idea is to propagate the initial conditions for a time $t$ from the set of all complex phase points $\{(\vec {\cal P}_\alpha, \vec {\cal Q}_\alpha)\}$ satisfying Eq.~(\ref{constraints}) and locate the intersections with the set $\{(\vec {\cal P}_\beta, \vec {\cal Q}_\beta)\}$ associated with the final state.  As each intersection indicates a saddle point, following the method of steepest descents gives a sum of contributions over these saddle points as the approximation to $\langle \beta | U_H(t)|\alpha\rangle$.  There are many possible complications, such as the existence of saddle points which must be thrown away, but this is a brief outline of the main ideas.  Carrying out the direct implementation of GGWPD means following this prescription, i.e.~solving the complex boundary value problem directly.

\subsection{Off-center, real trajectory sums}
\label{het}

Consider a system's classical transport and how its properties are captured by a time-dependent correlation function between an initially localized Gaussian density of phase space points $\rho_{\alpha}$ centered at point $(\vec p_\alpha, \vec q_\alpha)$ and a final destination Gaussian density $\rho_{\beta}$ centered at point $(\vec p_\beta, \vec q_\beta)$.  For simplicity, let $(\vec p_\alpha, \vec q_\alpha)$ and $(\vec p_\beta, \vec q_\beta)$ be periodic trajectories of a continuous dynamical system or fixed points of a dynamical map.   $t$ is the time of propagation.  In the limit of highly localized densities, it can be expressed as a sum over certain trajectories~\cite{Barnes93,Barnes94,Tomsovic91b,Oconnor92,Tomsovic93}
\begin{equation}\label{eq:one}
\Gamma_{\beta \alpha}(t) = \sum_{\gamma}\langle \rho_{\beta},T_\gamma^t \rho_{\alpha}\rangle
\end{equation}
where each $\gamma$ denotes a particular trajectory branch.  Each branch contains like-behaving trajectories that take time $t$ to leave the neighborhood of $(\vec p_\alpha, \vec q_\alpha)$ and arrive in the neighborhood of $(\vec p_\beta, \vec q_\beta)$.  Each term $\gamma$ represents a distinct dynamical pathway for connecting the two regions.  An individual $T^t_\gamma$ is a linearized dynamical time propagation representing the behavior of a branch labelled $\gamma$.  One trajectory is selected from $\gamma$ as its representative and its stability matrix is used to account for all the other members of that branch.  

The quantum mechanical analog in the semiclassical limit of the correlation function is
\begin{equation}
\label{eq:qm}
\left\langle \beta|U_H(t)| \alpha\right\rangle = \sum_{\gamma}\left\langle \beta|U_\gamma(t)| \alpha\right\rangle
\end{equation}
where $|\alpha\rangle$ is the ket vector corresponding to a quantum wave packet centered at $(\vec p_\alpha, \vec q_\alpha)$ and $U_\gamma (t)$ is the appropriately linearized unitary time translation operator corresponding to the canonical transformation represented by $T^t_\gamma$.  The summation is over the same branches as necessary for the classical expression.  These two equations hold equally well for open or bounded systems.  

There are Gaussian integrals to perform for these methods generated by considering one of the representative trajectories and making a quadratic expansion of its classical action to account for neighboring trajectories.  This quadratic expansion is effectively constructing a local, complex saddle, even though one is not seeking the location of its saddle point.  One is just doing the resultant Gaussian integrals.  The trajectories neighboring the representative one form a continuous set that create a path crossing this saddle.  This effective saddle would be the true saddle if the dynamics locally were strictly linear.  Local nonlinear dynamics introduce a degradation of the off-center, real trajectory method relative to GGWPD, but one sees the connection between the two methods; i.e.~$\gamma$ can be considered the index for real crossing saddles.    All of the parameters necessary to express those integrals' results analytically are given by the representative trajectory's classical action, geometric phase index, and stability matrix elements. A good review of the basic quadratic expansion techniques giving the explicit expressions relating the second action derivatives with stability matrix elements can be found in~\cite{Heller91}.

\subsubsection{Integrable dynamics}

Implementing off-center, real trajectory methods for integrable systems was developed in~\cite{Barnes93,Barnes94} and applied to the Coulomb potential with vanishing angular momentum.  The only way to classically transport from one region to the other is along tori that are common to both regions (intersect both regions).  Making a canonical transformation to local action-angle variables for the initial state region identifies the tori which potentially may intersect the final state region.  It is not necessary to run multiple trajectories on a single torus as that gives redundant transport information.  Instead, by fixing the angle variables and varying the actions, one is maximally exploring the set of transport possibilities.  Somewhat ideally, one would find the coordinate transformation to action-angle variables for the phase point $(\vec p_\alpha, \vec q_\alpha)$ and invoke the short wavelength approximation to justify expanding the Hamiltonian locally up to quadratic order in these action-angle variables; that gives the energy, and periods of motion and shearing rates for each of the degrees of freedom.  In practice, it is possible to avoid the explicit construction of local action-angle variables much like as done in~\cite{Barnes94}.  There approximate fixed angle trajectory manifolds are constructed intuitively for $(\vec p_\alpha, \vec q_\alpha)$ and $(\vec p_\beta, \vec q_\beta)$.  However, the method is very weakly dependent on the exact representative trajectory being used, and any of the other trajectories in its neighborhood give nearly the same result.  It is not really necessary to have the angle variables fixed as the action variables change over the manifold.  However selected, the former manifold is propagated a time $t$ numerically and intersections are found with the latter manifold.  

\subsubsection{Chaotic dynamics}

For chaotic systems, there are heteroclinic trajectories that converge to $(\vec p_\alpha, \vec q_\alpha)$ for $t\to -\infty$ and converge to $(\vec p_\beta, \vec q_\beta)$ for $t\to \infty$~\cite{Tomsovic91b,Oconnor92,Tomsovic93}, if $(\vec p_\alpha, \vec q_\alpha)$ and $(\vec p_\beta, \vec q_\beta)$ are chosen as periodic orbits.  Otherwise, the mental image becomes slightly more complicated.  They provide a convenient set of representative trajectories as they are in one-to-one correspondence with the necessary  $\gamma$ branches.  The summation is over finite-time segments of the infinite histories of the heteroclinic trajectories that satisfy the fixed time constraint.  This is a complete solution of the classical transport problem in the limit of a shrinking, localized density, which at some point is within a convergence zone shown to exist in the normal coordinate form~\cite{Moser56,Silva87}.  There are no transport pathways from the neighborhood of $(\vec p_\alpha, \vec q_\alpha)$ to the neighborhood of $(\vec p_\beta, \vec q_\beta)$ not accounted for by some heteroclinic trajectory.  The trajectories are found by constructing the unstable manifold of $(\vec p_\alpha, \vec q_\alpha)$ and the stable manifold of $(\vec p_\beta, \vec q_\beta)$, and intersecting them.  The necessary classical information and quadratic expansions follow by the same general methods as for the integrable systems.

\section{Implementing GGWPD}
\label{implement}

To begin, consider the essential ingredient in the method of steepest descents.  Once a complex trajectory that serves as a saddle point is found, there is a quadratic expansion performed about the saddle point, a deformation of a path for a line integral, and a Gaussian integral evaluated.  To the extent that the exponential argument is a purely quadratic function, one could expand it to second order about any other point and get exactly the same results as if expanded about the saddle point, i.e.~where the linear term disappears.  However, if there are cubic and higher order corrections, expanding about the saddle is optimal.  Depending on the strength of the higher order terms, there still has to be a neighborhood of points about the saddle point where if a quadratic expansion and integration are performed, the result will still be very nearly the same as the optimal one.  

Suppose that it is a real crossing saddle, i.e. there exists a path of real trajectories crossing it.  Suppose further that some of those trajectories are close enough to the true saddle trajectory to give nearly the same result; i.e.~both expansions lead to rather different-looking Gaussian integrals, but the implied saddle constructed by the quadratic expansion about the real trajectory has nearly the same shape and location.  Then one would also expect that one could construct a Newton-Raphson scheme starting with that real trajectory as an initial guess that would converge rapidly to the complex trajectory represented by the saddle point.  Indeed, the Newton-Raphson scheme has been introduced as a way of finding ``nearly-real'' complex trajectories~\cite{Vanvoorhis02, Vanvoorhis03}.  Here, the idea is to apply the Newton-Raphson scheme to the full set of representative, off-center, real trajectories for any given dynamical system.  In the case that the neglected local dynamical nonlinearities are strong enough, the Newton-Raphson scheme may not converge.  If so, some alteration of the search method must be incorporated, but there is still the $\gamma$ one-to-one correspondence of representative, off-center, real orbits and real crossing saddles up to the point where nonlinearities create saddles that coalesce and merge into each other.

As happens for any root in this scheme, there is a set of neighboring points in its neighborhood for which the method converges, which may be considered a basin of convergence.  Note that all the basins of convergence of various distinct roots must be mutually exclusive.  The representative, off-center, real trajectory lies somewhere on only one particular saddle, the one to which it is associated, even if it cannot be guaranteed that it is inside the basin of convergence.  If the representative trajectory is close enough and the local dynamics linear enough, the Newton-Raphson iterative procedure rapidly converges to the saddle point for that saddle.  The interpretation of the saddle point's contribution is given by its associated real trajectory branch.  

If the representative trajectory is not close enough or the branch has enough curvature, then before attempting to apply a Newton-Raphson search, it would be necessary to add an extra step in the process.  One might first find a more optimal real trajectory from the branch or perhaps, just to put a brake into the Newton-Raphson scheme.  Since this involves real-crossing saddles, there is a best trajectory from the branch to use (the one that lies on the curve of steepest "ascents").  After finding it, presumably the Newton-Raphson method would descend toward the actual saddle point.  Since the representative, off-center, real trajectory sums describe the complete classical transport, this method locates every saddle point that can be interpreted a result of a classically allowed process.  Those saddle points that are non-crossing must necessarily be related to classically non-allowed processes, i.e.~tunneling, or be on the wrong side of Stoke's surfaces and unphysical.  Such saddle points are left for future investigations.

It is known that for even the simplest functions with multiple roots of a single complex variable, the domains of convergence are fractals.  Nevertheless, exceptions to the one-to-one correspondence require sufficiently strong higher order curvature corrections to coalesce saddles.  The lowest order curvature correction is determined by the locally evaluated third derivatives of the action function.  These corrections turn out to be visible as strong curvatures in the branch under consideration.  So in spite of the complicated nature of the exact basins of convergence, it is clear when the search scheme from off-center, real trajectory to its associated saddle point may lead to failure.  

For continuous time dynamical systems, each saddle point found at some fixed time contributes throughout a continuous time window.  As time continuously changes, the saddle point trajectory continuously changes as well.  The trajectory also shifts its energy.  At some point, earlier in time and later in time, its energy is shifted so far outside of the energy uncertainty of the wave packet that its contribution fades away.  The end result of this one-dimensional parameter family of saddle points typically produces a chirped-like smooth function of time to propagating wave packets or correlation functions.  This is illustrated in the earlier work on the stadium in their Fig.~16~\cite{Tomsovic93}.

\subsection{Identifying the associated complex trajectories}

For the correlation function expression given in Eq.~(\ref{eq:qm}) consider one member from the set of off-center, real trajectory initial conditions, $\{\vec p_0^\gamma, \vec q_0^\gamma\}$.  Assuming the local dynamics are linear enough, it lies in the basin of attraction of a saddle point trajectory one wishes to locate.  Label its initial condition $\left( \vec{\cal{P}}_0^\gamma, \vec{\cal{Q}}_0^\gamma \right)$.  The constraints that the trajectory of interest must satisfy are obtained by subtracting the right hand side of Eq.~(\ref{constraints}) using the initial condition $\left( \vec{\cal{P}}_0^\gamma, \vec{\cal{Q}}_0^\gamma \right)$ that propagates to $\left( \vec{\cal{P}}_t ^\gamma, \vec{\cal{Q}}_t^\gamma \right)$.  Thus, a saddle point trajectory must satisfy
\begin{eqnarray}
2 {\bf b}_\alpha \cdot \left( \vec {\cal Q}_0^\gamma - \vec q_\alpha \right) + \frac{i}{\hbar} \left( \vec {\cal P}_0^\gamma -\vec p_\alpha\right )  &=&  \vec 0 \nonumber \\
2 {\bf b}_\beta \cdot \left( \vec {\cal Q}_t^\gamma - \vec q_\beta \right) - \frac{i}{\hbar} \left( \vec {\cal P}_t^\gamma -\vec p_\beta\right )  &=&  \vec 0
\label{GGWPDcondition}
\end{eqnarray}
where clearly this just indicates that $\left( \vec{\cal{P}}_0^\gamma, \vec{\cal{Q}}_0^\gamma \right)$ is one member of the set $\left\{\left( \vec{\cal{P}}_\alpha, \vec{\cal{Q}}_\alpha \right)\right\}$ and $\left(\vec{\cal{P}}_t^\gamma,  \vec{\cal{Q}}_t^\gamma \right)$ is the associated propagated member of the set $\left\{\left( \vec{\cal{P}}_\beta, \vec{\cal{Q}}_\beta \right)\right\}$.  

If some $\left(  \vec{\cal{P}}_0, \vec{\cal{Q}}_0 \right)$, such as one member of the set $\{ \vec p_0^\gamma, \vec q_0^\gamma\}$, is not a saddle point trajectory, then the right hand sides of Eqs.~(\ref{GGWPDcondition}) do not vanish, but rather equal some complex constant vectors $\vec{\cal{C}}_0$ and  $\vec{\cal{C}}_t$, respectively.  The basic idea is to use $(\vec{\cal{C}}_0,\vec{\cal{C}}_t)$ to solve for a shift of $\left( \vec{\cal{P}}_0, \vec{\cal{Q}}_0 \right)$ towards $\left( \vec{\cal{P}}_0^\gamma, \vec{\cal{Q}}_0^\gamma \right)$, i.e.~solve for
\begin{equation}
\begin{array}{l}
 \vec{\cal{P}}_0^\prime = \vec{\cal{P}}_0 + \delta \vec{\cal{P}}_0 \\
 \vec{\cal{Q}}_0^\prime = \vec{\cal{Q}}_0 + \delta\vec{\cal{Q}}_0
 \end{array}
\end{equation}
where $\left( \vec{\cal{P}}_0^\prime, \vec{\cal{Q}}_0^\prime \right)$ is a much better approximation of $\left( \vec{\cal{P}}_0^\gamma, \vec{\cal{Q}}_0^\gamma \right)$.  Then one updates the initial conditions $\left( \vec{\cal{P}}_0^\prime, \vec{\cal{Q}}_0^\prime \right) \rightarrow \left( \vec{\cal{P}}_0, \vec{\cal{Q}}_0 \right)$, recalculates $(\vec{\cal{C}}_0,\vec{\cal{C}}_t)$, and solves for a new set of shifts $\left(  \delta \vec{\cal{P}}_0, \delta \vec{\cal{Q}}_0 \right)$.  This is repeated until the procedure converges to the saddle point of interest.

In Eq.~(\ref{GGWPDcondition}) there appears to be $8D$ unknown quantities and only $4D$ constraints, but $\left( \vec{\cal{P}}_t ^\gamma, \vec{\cal{Q}}_t^\gamma \right)$ is precisely determined by the initial conditions $\left( \vec{\cal{P}}_0^\gamma, \vec{\cal{Q}}_0^\gamma \right)$.  Thus, the shifts $\left( \delta \vec{\cal{P}}_0, \delta \vec{\cal{Q}}_0 \right)$ also determine $\left( \delta \vec{\cal{P}}_t, \delta \vec{\cal{Q}}_t \right)$.  For small $\left( \delta \vec{\cal{P}}_0, \delta \vec{\cal{Q}}_0 \right)$, a very accurate approximation is to use the stability matrix of the trajectory whose initial conditions are given by $\left( \vec{\cal{P}}_0, \vec{\cal{Q}}_0 \right)$,
\begin{equation}
\left( \begin{array}{c} \delta \vec{\cal{P}}_t \\ \delta \vec{\cal{Q}}_t \end{array} \right) =  \left( \begin{array}{c} \bf{M_{11}} \\ \bf{M_{21}} \end{array} \begin{array}{c} \bf{M_{12}} \\ \bf{M_{22}} \end{array} \right)
\left( \begin{array}{c} \delta \vec{\cal{P}}_0 \\ \delta \vec{\cal{Q}}_0 \end{array} \right) 
\label{delta}
\end{equation}
to replace the quantities $\left( \delta \vec{\cal{P}}_t, \delta \vec{\cal{Q}}_t \right)$.  This approximation is the complex, multi-dimensional equivalent of using the slope in a Newton-Raphson recursive root finding method.  After a couple steps of algebra, the relations to use recursively are found to be,
\begin{eqnarray}
 -\vec{\cal{C}}_0  &=& 2 {\bf b}_\alpha \cdot \delta \vec{\cal{Q}}_0 + \frac{i}{\hbar} \delta \vec{\cal{P}}_0 \nonumber \\
 - \vec{\cal{C}}_t &=&  2 {\bf b}_\beta \cdot \left( {\bf{M_{21}}} \cdot \delta \vec{\cal{P}}_0+{\bf{M_{22}}} \cdot \delta \vec{\cal{Q}}_0 \right) - \nonumber \\
 && \frac{i}{\hbar} \left( {\bf{M_{11}}} \cdot \delta \vec{\cal{P}}_0 + {\bf{M_{12}} } \cdot \delta \vec{\cal{Q}}_0\right)
\label{GGWPDcorrection2}
\end{eqnarray}
Being linear, these equations are straightforwardly solved for $\left( \delta\vec{\cal{P}}_0, \delta\vec{\cal{Q}}_0 \right)$.
The updated constants $\vec{\cal{C}}_0$ and  $\vec{\cal{C}}_t$ rapidly approach null vectors, and thus the sequence of updated $\left( \vec{\cal{P}}_0, \vec{\cal{Q}}_0 \right)$ similarly approach the saddle point trajectory $\left( \vec{\cal{P}}_0^\gamma, \vec{\cal{Q}}_0^\gamma \right)$ associated with the classically allowed transport following $( \vec p_0^\gamma, \vec q_0^\gamma )$.

\subsection{Explicit saddle point expressions}
\label{expressions}

The explicit multi-degree-of-freedom expressions deriving from the GGWPD method can be determined using an analytically continued version of the van Vleck-Gutzwiller propagator for incorporating the quantum dynamics; see~\cite{Novaes05}.  Attention must be paid to both the phase and prefactor.  A quadratic expansion of the exponential argument in terms of stability matrix elements gives the most explicit form of the results.

\subsubsection{Taylor series expansion of the propagator}

The usual form of the van Vleck-Gutzwiller propagator is given as
\begin{eqnarray}
\label{vanvleck}
G(\vec x, \vec x^\prime; t) &=& \left( \frac{1}{2\pi i\hbar} \right)^{D/2} \sum_\gamma \left| \frac{1}{{\rm Det\left({\bf M}_{21}\right)_\gamma}} \right|^{1/2} \qquad \qquad \nonumber \\
&& \qquad \quad \times \exp \left( \frac{i}{\hbar} S_\gamma (\vec x, \vec x^\prime; t) - \frac{i\nu_\gamma \pi}{2}\right)
\end{eqnarray}
where the $\gamma$ summation is over all the trajectories that begin at $\vec x^\prime$ and finish at $\vec x$ in exactly a time $t$.  In order to use this propagator to arrive at the desired results, there are two main issues that need to be addressed, the multi-dimensional complex quadratic expansion of the action function and how to interpret the absolute value of the determinant after analytical continuation to complex variables.

One way to obtain the desired relationship between second derivatives of a complexified version of $S_\gamma (\vec x, \vec x^\prime; t)$ and stability matrix elements follows by recognizing that the relation
\begin{equation}
\left(\begin{array}{c} \delta \vec p_t \\ - \delta  \vec p_0 \end{array}\right) = \left.
\left(\begin{array}{cc}  \frac{\partial^2 S_\gamma }{\partial \vec x \partial \vec x} & \frac{\partial^2 S_\gamma }{\partial \vec x^\prime \partial \vec x} \\ \\ \frac{\partial^2 S_\gamma }{\partial \vec x \partial \vec x^\prime} & \frac{\partial^2 S_\gamma }{\partial \vec x^\prime \partial \vec x^\prime} \end{array}\right)\right|_{\vec q_t,\vec q_0} \cdot
\left(\begin{array}{c}  \delta \vec q_t \\ \delta \vec q_0 \end{array}\right)   
\end{equation}
can be algebraically rearranged for the variables to be in the same column vectors as found in Eq.~(\ref{delta}), and it is sufficient to reinterpret the derivatives as being due to complex positions.  This algebraic rearrangement generates the relations
\begin{widetext}
\begin{equation}
\left(\begin{array}{cc}  \frac{\partial^2 S_\gamma }{\partial \vec{\cal{Q}}_t \partial \vec{\cal{Q}}_t} & \frac{\partial^2 S_\gamma }{\partial \vec{\cal{Q}}_0 \partial \vec{\cal{Q}}_t} \\ \\ \frac{\partial^2 S_\gamma }{\partial \vec{\cal{Q}}_t \partial \vec{\cal{Q}}_0} & \frac{\partial^2 S_\gamma }{\partial \vec{\cal{Q}}_0 \partial \vec{\cal{Q}}_0} \end{array}\right) = 
\left( \begin{array}{cc} \bf{M_{11}}\cdot  \bf{M_{21}}^{-1} & \quad \bf{M_{12}} - \bf{M_{11}} \cdot  \bf{M_{21}}^{-1} \cdot  \bf{M_{22}} \\ 
- \bf{M_{21}}^{-1} & \bf{M_{21}}^{-1} \cdot \bf{M_{22}} \end{array} \right)
\end{equation}
\end{widetext}
where the stability matrix elements may now be complex.  Thinking ahead of using this matrix for evaluating multivariate Gaussian integrals, it is necessary to know whether it is symmetric.  That is assured by the possibility of applying the derivatives in the opposite order and obtaining the same result.  It is amusing to note that for free particle propagation of complex initial conditions, the stability elements remain real, but for general dynamical systems that is not the case.  

Next consider the question about the interpretation of the absolute value of the determinant.  For real trajectories, it can only be positive or negative.  Taking the absolute value and using the geometric index $\nu$ is just a way of keeping track of the sheet of a multivalued function, and making sure that phase variation remains smooth (no discontinuous jumps when changing sheets of the function).  Extended to complex trajectories, the determinant's phase can be anywhere on the unit circle, not just plus or minus unity.  However, applying the same logic of maintaining smooth phase evolution, the phase index is just keeping track of the correct square root branch (it has a positive real part).  One does not want the absolute value, the phase is necessary as well. Thus, the issue is resolved by taking the correct branch of the square root, not taking the absolute value.  The absolute value notation is dropped on the determinant, but no new notation is added to indicate taking the correct branch of the square root.  It is left implied.  Ahead, this enables algebraic simplifications involving this square root, and it eventually disappears. \\

\subsubsection{Time-evolving wave functions}

All the ingredients exist at this point to write down the explicit expression for the full GGWPD multi-degree-of-freedom equation for the evolution of the wave packet $\phi_\alpha (\vec x)$.  It is given by applying Eq.~(\ref{handy}) and doing a bit of algebra,
\begin{widetext}
\begin{eqnarray}
\label{psit}
\phi_\alpha (\vec x;t) &=& \int_{-\infty}^\infty {\rm d}\vec x^\prime G(\vec x, \vec x^\prime; t) \phi_\alpha (\vec x^\prime) \nonumber \\
&=& \left( \frac{1}{2\pi i\hbar} \right)^{D/2} \left(\frac{2^D{\rm Det}[{\bf b}_\alpha]}{\pi^D}\right)^{1/4} \sum_\gamma \exp \left[ \frac{i}{\hbar} S(\vec x,\vec {\cal Q}_0^\gamma;t) - \frac{i\nu_\gamma \pi}{2} + F^-_0(\alpha,\gamma) \right]  \left( \frac{1}{{\rm Det\left[{\bf M}_{21}\right]_\gamma}} \right)^{1/2} \nonumber \\
&& \int_{-\infty}^\infty {\rm d}\vec x^\prime \exp \left[ - \left(\vec x^\prime - \vec {\cal Q}_0^\gamma \right) \cdot \left( {\bf b}_\alpha - \frac{i}{2\hbar}{\bf M}^{-1}_{21} {\bf M}_{22} \right) \cdot \left(\vec x^\prime - \vec {\cal Q}_0^\gamma \right) \right] \nonumber \\
&=& \left( \frac{2^D{\rm Det}\left[ {\bf b}_\alpha \right]}{\pi^D} \right)^{1/4} \sum_\gamma \frac{\exp \left[  \frac{i}{\hbar} S(\vec x,\vec {\cal Q}_0^\gamma;t) - \frac{i\nu_\gamma \pi}{2} + F^-_0 (\alpha,\gamma) \right]}{\left( {\rm Det}\left[ {\bf M}_{22} +2i\hbar {\bf M}_{21} \cdot {\bf b}_\alpha \right] \right)^{1/2}} \nonumber \\
\end{eqnarray}
\end{widetext}
The set of saddle point trajectories $\gamma$ depends in a complicated way on the position $\vec x$ and the dynamics generated by $H$, and thus $\phi_\alpha (\vec x)$ is no longer in the form of a Gaussian wave packet nor a sum over Gaussian wave packets.  However, if $H$ is no greater than quadratic in the phase space variables, the sum reduces to a single term.  The expression can, after some work, be reduced to that of linearized wave packet dynamics, which of course, gives a Gaussian wave packet form.  Because the saddle point depends continuously on the value of $\vec x$ and there is only one trajectory for linearized wave packet dynamics, the work involved is to show that for every value $\vec x$, the two expressions are equivalent.

\subsubsection{Correlation functions}

Similarly, all the ingredients exist to give the explicit expression for time-dependent correlation functions $\langle \beta | U_H(t) |\alpha \rangle$, or less generally, the coherent state representation of the time-dependent Green function.  The necessary algebra is slightly more complicated and three linear algebra identities are very helpful.  They are in no particular order,
\begin{eqnarray}
{\rm Det}[{\bf A}] {\rm Det}[{\bf B}]&=& {\rm Det}\left[ {\bf A} {\bf B} \right]  \nonumber \\
{\rm Det}[{\bf A}] &=& {\rm Det}\left[ \begin{array}{cc} \mathbf{1} & \mathbf{0} \\ \mathbf{0} & \mathbf{A} \end{array} \right] \nonumber \\
{\rm Det}\left[ \begin{array}{cc} \mathbf{A} & \mathbf{B} \\ \mathbf{C} & \mathbf{D} \end{array} \right] &=& {\rm Det}\left[ \mathbf{A} - \mathbf{B} \mathbf{D}^{-1} \mathbf{C} \right] {\rm Det} \left[ {\bf D} \right] 
\end{eqnarray}
After  applying Eq.~(\ref{handy}) and some linear algebra 
\begin{widetext}
\begin{eqnarray}
\label{corrt}
\langle \beta | U_H(t) |\alpha \rangle &=& \int_{-\infty}^\infty {\rm d}\vec x {\rm d}\vec x^\prime \phi_\beta^* (\vec x) G(\vec x, \vec x^\prime; t) \phi_\alpha (\vec x^\prime) \nonumber \\
&=& \left( \frac{1}{2\pi i\hbar} \right)^\frac{D}{2} {\cal N}_\alpha^0 {\cal N}_\beta^t  \sum_\gamma \exp \left[ \frac{i}{\hbar} S(\vec {\cal Q}_t^\gamma,\vec {\cal Q}_0^\gamma;t) - \frac{i\nu_\gamma \pi}{2}  \right]  \left( \frac{1}{{\rm Det\left[{\bf M}_{21}\right]_\gamma}} \right)^\frac{1}{2}  \int_{-\infty}^\infty {\rm d}\vec x {\rm d}\vec x^\prime\nonumber \\
&&  \exp \left[ - \left(\vec x - \vec {\cal Q}_t^\gamma , \vec x^\prime - \vec {\cal Q}_0^\gamma \right) \cdot \left[ \begin{array}{cc} {\bf b}_\beta - \frac{i}{2\hbar}{\bf M}_{11} {\bf M}^{-1}_{21} & \ \   \frac{i}{2\hbar}\left( {\bf M}_{11} \cdot {\bf M}^{-1}_{21} \cdot {\bf M}_{22} - {\bf M}_{12} \right) \\ \\ 
\frac{i}{2\hbar}{\bf M}^{-1}_{21} & {\bf b}_\alpha - \frac{i}{2\hbar}{\bf M}^{-1}_{21} {\bf M}_{22} \end{array} \right] \cdot \left(\begin{array}{c} \vec x - \vec {\cal Q}_t^\gamma\\ \vec x^\prime - \vec {\cal Q}_0^\gamma \end{array} \right) \right] \nonumber \\ \nonumber \\
&=& \left( 4^D{\rm Det}\left[ {\bf b}_\alpha \right] {\rm Det}[{\bf b}_\beta]\right)^{1/4}  \sum_\gamma \frac{\exp \left[  \frac{i}{\hbar} S(\vec {\cal Q}_t^\gamma,\vec {\cal Q}_0^\gamma;t) - \frac{ i\nu_\gamma \pi}{2} + F^-_0  (\alpha,\gamma) + F^+_t (\beta,\gamma) \right]}{\left( {\rm Det}\left[{\bf M_{11} \cdot b_\alpha } + {\bf b_\beta \cdot M}_{22} +2i\hbar {\bf b_\beta \cdot M_{21} \cdot b_\alpha} - \frac{i}{2\hbar}{\bf M_{12}} \right] \right)^{1/2}} 
\end{eqnarray}
\end{widetext}
It is worth noting that this expression, at least superficially, is much simpler than the one required for off-center, real trajectory methods because of the complicated looking terms introduced by non-vanishing linear terms in the Gaussian integrals; see the Appendix.  It is also separated into parts that can be interpreted more easily.  There is the classical action of the complex trajectory, its stability matrix, and the functions $F^-_0  (\alpha,\gamma)$ and $F^+_t (\beta,\gamma)$ that contain a phase and a measure of how complex are the initial and final coordinates of the trajectory.  This expression is used with the kicked rotor in the next section to illustrate the comparison of accuracy of GGWPD and off-center, real trajectory methods.

\subsubsection{Comparing semiclassical methods: instructive analytic example}

It is instructive to compare the basic workings of linearized wave packet dynamics, off-center, real trajectory methods, and GGWPD, where this ordering is in terms of increasing sophistication.  Perhaps, the simplest example in which the trajectories can be worked out analytically for all three approximations is for free particle propagation.  In addition, all three approximations are exact, and thus contain identical information.  They are just obtaining it from the trajectories and organizing it in their respective expressions differently.  

For the purposes of this comparison, a single degree of freedom gives a sufficient illustration. Let ${\bf b}^{-1}_\alpha = 4\sigma^2$ and $\kappa = \hbar t/(2 m \sigma^2)$.  The stability matrix for every possible trajectory, real or complex is
\begin{equation}
{\bf M}_t = \left(\begin{array}{cc} 1 & 0 \\ t/m & 1 \end{array}\right)
\end{equation}
The exact expression for a propagated wave packet can be written in the form that comes from linearized wave packet dynamics as follows
\begin{eqnarray}
\label{exact}
\phi_\alpha (\vec x;t) &=& \left(\frac{1}{2\pi\sigma^2}\right)^{1/4} \left(\frac{1}{1+i\kappa}\right)^{1/2} \times \nonumber \\
&& \exp \left[ -\frac{\left(x-q_t\right)^2}{4\sigma^2 (1+i\kappa)}  + \frac{ip_t}{\hbar}\left(x-q_t\right) + \frac{ip_t^2 t}{2 m \hbar}\right] \nonumber \\
\end{eqnarray}
where the trajectory with initial conditions $(p_0, q_0)=(p_\alpha, q_\alpha)$ gives the classical trajectory
\begin{equation}
\begin{array}{cr} p_t & = \\ q_t & = \end{array}  \begin{array}{l} p_\alpha  \\ q_\alpha + \frac{t}{m}p_\alpha \end{array} 
\end{equation}
The last term in the exponential is the phase that comes from $\int^t {\cal L} {\rm d}t^\prime$, $\kappa$ ``chirps'' the Gaussian, i.e.~puts fast phase oscillation out front and slow phase oscillation behind the wave packet center, and $(p_t, q_t)$ shifts the center appropriately.

The off-center, real trajectory method arrives at this result in a very different way.  A different trajectory is used for every argument $x$ of the wave packet.  Given that momentum is the action variable for free particle motion, to locate the representative off-center, real trajectory for some given value of $x$, one propagates the set of initial conditions $(p_0, q_\alpha)$ for all real $p_0$, and finds which one intersects the with the set $(p_t, x)$ for all $p_t$.  The initial conditions and trajectories as a function of $x$ are thus,
\begin{equation}
\begin{array}{cr} p_0 & = \\ q_0 & = \end{array}  \begin{array}{l} p_\alpha + \frac{m}{t}\left(x-q_t\right) \\ q_\alpha   \end{array}
\end{equation}
Only if $x$ coincides with the moving wave packet's center does the off-center, real trajectory equal the linearized wave packet trajectory.  Otherwise, the initial momentum is detuned from $p_\alpha$ so that the trajectory arrives exactly at $x$ no matter what value of $x$ or time $t$ is considered.  As time varies, the method uses trajectories across the entire wave packet, short times - high momenta, long times - low momenta.  In more general nonlinear dynamical systems, the local dynamics of all the phase space of the packet is explored and there is the capacity to include multiple contributions that cannot be accounted for in linearized wave packet dynamics.   The improvements with this method for problems with nonlinear dynamics are optimized if the ${\bf b}_\alpha$ matrix stretches the wave packet more along the set of propagated initial conditions rather than transverse.

The GGWPD method improves further on the off-center, real trajectory method by incorporating information about the shape of the initial wave packet into the complexified trajectories.  The initial conditions for the saddle point trajectories are given by
\begin{eqnarray}
{\cal P}_0 & = & p_\alpha + \frac{i\kappa m}{t}\left(\frac{x-q_t}{1+i\kappa}  \right) \nonumber \\
{\cal Q}_0 & = & q_\alpha + \frac{x-q_t}{1+i \kappa}
\end{eqnarray}
The initial conditions are a function of the final real position $x$ as for the off-center, real trajectory method, but are shifted from the linearized wave packet trajectory by complex terms that contain information about the shearing and shape parameter $\kappa$, and the distance of $x$ from the moving wave packet center $q_t$.  

Since the off-center, real trajectories are supposed to be close to the saddle point trajectories, consider the comparison of the real parts of the initial conditions of the saddle point trajectories and the off-center, real trajectory initial conditions.  After some algebra, one finds
\begin{eqnarray}
{\cal P}_0^R - p_0 &=& - \frac{m}{t} \left(\frac{x-q_t }{1+\kappa^2} \right) \nonumber \\
{\cal Q}_0^R - q_0 &=&  \frac{x-q_t}{1+\kappa^2} 
 \end{eqnarray} 
The wave packet amplitude is extremely small unless $x$ is within a few widths of $q_t$, and $\kappa$ increases proportionally with time.  Thus, for $x$ where there is some chance of non-negligible contribution, the real parts of the saddle point trajectories approach the off-center, real trajectory initial conditions as $t$ increases.  Consistent with this, the imaginary parts shrink with time as well.  However, it can be rather misleading to think of the off-center, real trajectory as being near the saddle point trajectory in any case where $x$ is not so close to $q_t$.  Finally, it turns out that the final term in $F^-_0(\alpha)$ from Eq.~(\ref{norm2}) (has real momentum and imaginary position) cancels with the damping term in the complexified classical action.  Thus, for free particle motion, the overall damping factor is given by the other two real terms in that expression, which involve only the squared imaginary parts of the position and momentum, respectively.

Overall, this simple example helps to illustrate the increasing level of sophistication of the three kinds of semiclassical approximations.  The first, linearized wave packet dynamics, is fine until nonlinear dynamics appears.  This is the Ehrenfest time scale, beyond which, being restricted to a single orbit, it has no built-in mechanism to handle the nonlinearities.  The off-center, real trajectory method builds in all the nonlinearities in the dynamics and is not limited by the Ehrenfest time scale whatsoever, but the nonlinearities are accounted for without taking account of the wave packet shape.  GGWPD does do this, and in addition, though beyond the scope of this paper, has the capacity to incorporate tunneling through the existence of additional saddle points.  So long as transport is dominant (tunneling is a tiny component) and the wave packet shape is well adapted to the problem, the off-center, real trajectory method and GGWPD should return compatible results.  In the next section, GGWPD is seen to be more accurate though.

\section{The kicked rotor}
\label{rotor}

The kicked rotor is a simple, yet extraordinary paradigm for both classical and quantum dynamical systems, which as a function of a parameter, spans the possibilities from classically integrable to strongly chaotic dynamics.  It has also been experimentally realized with cold atoms and a BEC in a kicked optical lattice~\cite{Moore95, Ryu06}.  A great deal is known about its classical and quantum dynamics~\cite{Chirikov79,Izrailev90,Lakshminarayan97}.  It is a mechanical-type particle constrained to move on a ring that is kicked instantaneously every multiple of a unit time, $t=n$.  The Hamiltonian is 
\begin{equation}
\label{krg}
H(q,p) = \frac{p^2}{2} - \frac{K}{4\pi^2}\cos (2\pi q) \sum_{n=-\infty}^\infty \delta(t-n) 
\end{equation}
The classical mapping equations for the version on the unit phase space torus are:
\begin{equation} 
\label{eq:two}
\begin{split}
& p_{n+1} =p_{n}-\frac{K}{2\pi }\sin (2\pi q_{n}) \pmod 1 \\
& q_{n+1} =q_{n}+p_{n+1} \pmod 1
\end{split}
\end{equation} 
For the kicking strength parameter $K=0$, the system is integrable, and for very small values, the system remains nearly integrable.  For $K$ values exceeding $2\pi$, the system is strongly and almost completely chaotic.  The stability matrix for a single iteration of a trajectory is,
\be
{\bf M}_n = \l( \barr{cc} 1 & -K \cos \l(2 \pi q_n \r) \\ 1 & 1-K \cos \l(2 \pi q_n \r) \earr \r ) \;,
\label{deltas1}
\ee
which are multiplied together consecutively for trajectories with greater numbers of iterations.

In its quantum realization, the dynamics are generated by iterations of the unitary Floquet operator,
\begin{equation}
\widehat{F} = \exp\l( \dis\frac{-i \widehat{p}^2}{2\hbar} \r) \; \exp\l[ \dis\frac{iK}{4\pi\hbar^2}  \cos 2\pi\widehat{q} \r] \;.
\label{eq.4a3}
\end{equation}
Its corresponding matrix elements in configuration space, $F_{rs} = \langle q_r \mid \widehat{F} \mid q_s \rangle$, are
\begin{equation}
\resizebox{.9\hsize}{!}{$F_{rs} = \dis\frac{1}{\dis\sqrt{iN}} \exp\l[ \dis\frac{i\pi (r-s)^2}{N} \r] \exp\l[ \dis\frac{iNK}{2\pi} \cos\dis\frac{2\pi s}{N} \r]$ \;,}
\label{eq.4a4}
\end{equation}
where $N$ is the Hilbert space dimension; $1 \leq r,s \leq N$. Thus, the value of Planck's constant is fixed by $2\pi\hbar N = 1$.  By increasing $N$, or equivalently decreasing $\hbar$, the semiclassical limit can be studied in great detail.  The initial state is taken to be a wave packet whose position representation is evaluated at the discrete set of $N$ position values $q_s = s/N$ using Eq.~(\ref{wavepacket}) to obtain the amplitude with the caveat that due to the discrete nature of the basis, the prefactor is somewhat modified in order to obtain a true unit norm.  Propagation then proceeds by the expected matrix multiplication with the Floquet operator.  

In all of the calculations, the choice of position representation variance of the wave packet ($b_\alpha^{-1} = 4 \sigma^2 = 2\hbar = 1/(\pi N)$) is such that the momentum and position uncertainties are equal.  Thus, the shape of the wave packet's phase space analogy appears circular in the plots and remains so for all values of $\hbar$.  The area inside the $2\sigma$ contour for any wave packet then is equal to $h=1/N$.  As mentioned above, the saddle points are independent of $\hbar$ with this choice, which simplifies the interpretation and calculations for comparing the accuracies of the semiclassical approximations as a function of $\hbar$.

\subsection{Preliminaries}
\label{prelim}

The geometry of phase space is more complicated for the phase space with complex ($ \cal \vec P, \vec Q$), and is no longer a torus as the space is not periodic in imaginary position and momentum.  For this and other reasons, in the phase space of real variables, it turns out to be convenient to use the ``unfolded torus", meaning that by not invoking the modulus $1$ operations in the mapping equations, there is a ``flat" phase space that extends to infinity.  Each unit square is a repetition of the fundamental torus, which is taken to be the $[0,1) \times [0,1)$ square.  Any two real points separated by an integer in either $q$ or $p$ are the same point; i.e.~$(p,q)$ and $(p+n_{p}, q+n_{q})$ are the same point.  The integers $n_{p}$ and $n_{q}$ can be thought of as winding numbers (including negative integers), i.e.~how many times a particle has wrapped around the cycles of the torus, which generally do have phase consequences in quantum mechanics.  A Gaussian wave packet has an image within each unit square.  However, for small enough values of $\hbar$ and our choice of variance, the tails of the images have fallen sufficiently in magnitude before reaching the original wave packet, and can be neglected.  

  It is sufficient to understand the classical transport of the single Gaussian density for the purposes of finding the set $\{\gamma\}$ of trajectories to use for implementing GGWPD; one must consider its contributions to all the final images in determining $\{\gamma\}$ however.  The analytically continued mapping equations, stability matrices, and action functions are needed for this.  They are given by
\begin{eqnarray}
{\cal P}_{n+1} &=& {\cal P}_{n} - \frac{K}{2\pi }\sin \left(2\pi {\cal Q}_{n} \right) \nonumber \\
{\cal Q}_{n+1} &=& {\cal Q}_{n} + {\cal P}_{n+1} \nonumber \\
{\bf M}_n &=& \left( \begin{array}{cc} 1 & - K \cos \left(2 \pi {\cal Q}_n \right) \\ 1 & 1-K \cos \left(2 \pi {\cal Q}_n \right) \end{array} \right) \nonumber \\
S\left({\cal Q}_t, {\cal Q}_0\right) &=& \sum_{n=0}^{t-1} S\left({\cal Q}_{n+1}, {\cal Q}_n\right) \nonumber \\
S\left({\cal Q}_{n+1}, {\cal Q}_n\right) &=& \frac{\left({\cal Q}_{n+1}-{\cal Q}_n\right)^2}{2} + \frac{K}{4\pi^2} \cos \left(2\pi {\cal Q}_n \right) \nonumber \\
\end{eqnarray}
where the modulus operation is not being applied.

\subsection{An integrable example}
\label{res}

The goals of providing the kicked rotor examples are to demonstrate: i) the relationship between the representative, off-center, real trajectories and the saddle points for both integrable and chaotic systems, ii) the improvements that arise using the full GGWPD, particularly as compared with off-center, real trajectory methods, and iii) the behavior in the semiclassical limit, i.e.~$\hbar \rightarrow 0$.  For this it is helpful to identify an example with at most one or two saddle points.  To begin, consider an integrable (near-integrable) system for which $K=0.05$.  The real phase space structure, and initial and final classical 
\begin{figure}[htb]
\includegraphics[width=3.45in]{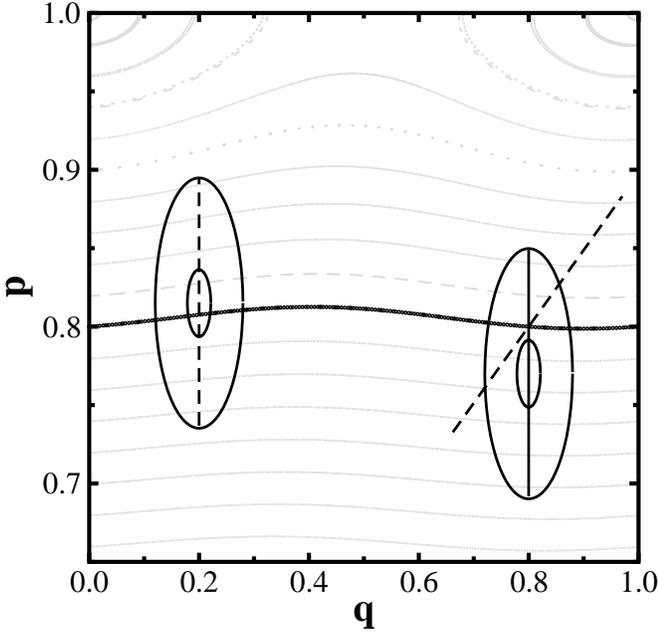}
\caption{The phase space structure of the example integrable dynamics for which $K=0.05$.  An initial classical Gaussian density is centered at $(p_\alpha, q_\alpha)=(0.815,0.2)$ and a final one at $(p_\beta, q_\beta)=(0.77,0.8)$.  The $2\sigma$ contours are shown that correspond to the large and small limiting values of $\hbar$ in the quantum calculations.  The light, mostly horizontal curves are the tori underlying the dynamics.  The sloped dashed line on the right is the result of propagating the initially vertical dashed line of initial conditions pictured on the left for two time steps.  The off-center, real trajectory lies at the intersection of the dashed sloped and solid vertical lines on the right.  The torus on which it resides is darkened.}
\label{fig1}
\end{figure}
densities corresponding to Gaussian wave packets are shown in Fig.~\ref{fig1}.  The initial wave packet's analogous classical density is pictured on the left.  The circles represent the $2\sigma$ contours for the density.  The large circle is for $h=1/50$ and the small one for $h=1/700$, the range of $h$ over which the $\hbar$-dependence is calculated in the various figures.  The same is pictured on the right for the final wave packet's analogous classical density.  The vertical line through the center of the left density is the collection of shearing trajectories that the off-center, real trajectory method uses.  After two iterations forward in time, it ends up as the long sloped line.  Its intersection with the right vertical line representing the final state's trajectory collection gives the off-center, real trajectory to be used in the off-center, real trajectory method~\cite{Barnes93,Barnes94}.  It also provides the initial guess for finding the true saddle point.  In this example, there is clearly one and only one saddle point.  The fact that the sloped line is so straight indicates that locally the dynamics are very well captured by a linearization, and that the off-center, real trajectory method is expected to give an excellent approximation.  One also expects the Newton-Raphson method to rapidly converge under such circumstances and the saddle point trajectory is converged to double precision with four iterations.  Note that the off-center, real trajectory method is better optimized than linearized wave packet dynamics, and a better approximation. 

\begin{figure}[htb]
\includegraphics[width=3.45in]{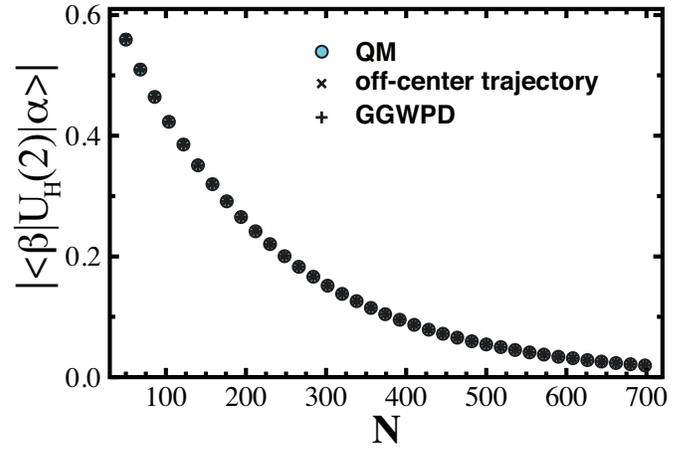}
\caption{ The absolute magnitude of the quantum, off-center, real trajectory, and GGWPD results versus $N$.  The off-center, real trajectory initial condition is $(p_0,q_0)=(0.8075799,0.20)$ and the saddle point $({\cal P}_0, {\cal Q}_0) = (0.8019843+i0.0062830, 0.2062830+i0.0130157)$.  Both semiclassical methods give very accurate results.  The overall decrease with increasing $N$ is due to the fact that the overlapping density gets further away from the wave packet centers measured in terms of the shrinking widths.}
\label{fig2}
\end{figure}

The overall accuracy of the two semiclassical methods are compared in Fig.~\ref{fig2}.  The overlap magnitude of 
$\langle \beta | U_H(t=2) |\alpha \rangle$ is plotted as a function of inverse-$\hbar$ ($N$).  Clearly, both semiclassical
theories accurately represent the quantum propagation.  It is not possible to see which semiclassical theory is the better approximation from this figure.  A more sensitive view of the absolute magnitude of the errors in the semiclassical approximations is shown in 
\begin{figure}[htb]
\includegraphics[width=3.45in]{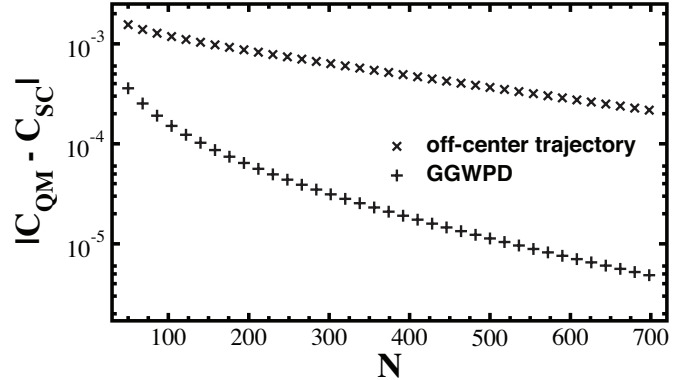}
\caption{Differences of the quantum overlap and semiclassical approximations.  The notation on the $y$-axis is $C= \langle \beta | U_H(2) |\alpha \rangle$ with the subscript indicating the evaluation. The difference of the magnitudes are shown on a logarithmic scale. }
\label{fig3}
\end{figure}
Fig.~\ref{fig3}.  The agreement is excellent for both semiclassical methods, but GGWPD is improving more quickly as $\hbar\rightarrow 0$ and is approaching two orders of magnitude of improvement.  Given how linear the shearing trajectory manifold slicing through final wave packet is, this is a somewhat surprising improvement in accuracy.  

\begin{figure}[htb]
\includegraphics[width=3.45in]{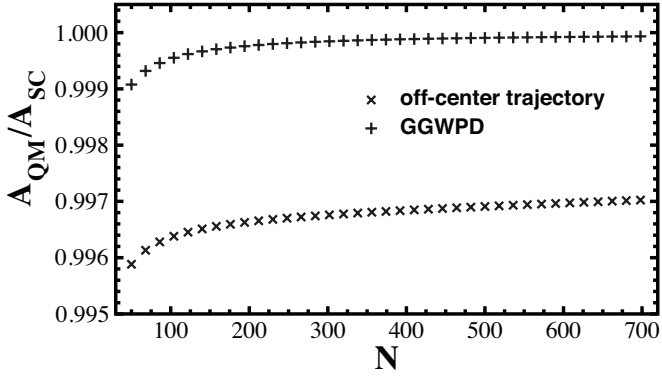}
\caption{Relative magnitude errors of the two semiclassical methods.  The notation on the $y$-axis is $A= |\langle \beta | U_H(2) |\alpha \rangle |$ with the subscript indicating the evaluation. The ratio of the quantum magnitude to the semiclassical magnitudes are shown as a function of $N$.  Note that whereas the GGWPD method approaches unity with increasing $N$, the off-center, real trajectory method apparently does not.}
\label{fig4}
\end{figure}

\begin{figure}[htb]
\includegraphics[width=3.3in]{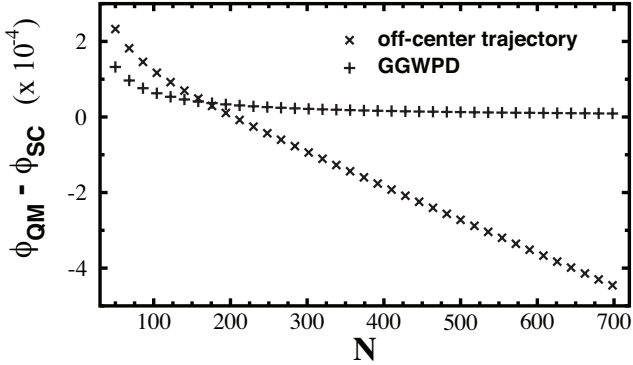}
\caption{ Phase errors of the two semiclassical methods.  The notation is $A{\rm e}^{i\phi}= \langle \beta | U_H(2) |\alpha \rangle$ with the subscript indicating the method.}
\label{fig5}
\end{figure}

There are two interesting features of the errors in the off-center, real trajectory method not present in the GGWPD results.  First as shown in Fig.~\ref{fig4}, although the absolute error is decreasing, the relative error does not appear to vanishing for the off-center, real trajectory method.  This is seen by the fact that the ratio of quantum magnitude to its magnitude does not appear to be approaching unity as it does for GGWPD.  So it has the feature of an absolute error which is shrinking even though its relative error is not.  There is no contradiction with the idea that in the $\hbar$-limit the 
error should vanish because in this limit the off-center, real trajectory moves further and further away from the wave packet center measured in widths.  At some point, it is beyond or outside the phase space region that needs to be taken into account and it would not even be included.  If at longer times there were a trajectory close enough to the wave packet centers to be included, it would give a more accurate contribution.  Second as seen in Fig.~\ref{fig5}, there is a 
slight drift in the phase relative to the quantum phase.  It is rather small, but unlike in GGWPD, it does not approach a 
vanishing phase difference.  In spite of these two features, the off-center, real trajectory method is nevertheless quite good.  It is just not as excellent as GGWPD.  

\subsection{A chaotic example}

\begin{figure}[htb!]
\includegraphics[width=2.4in]{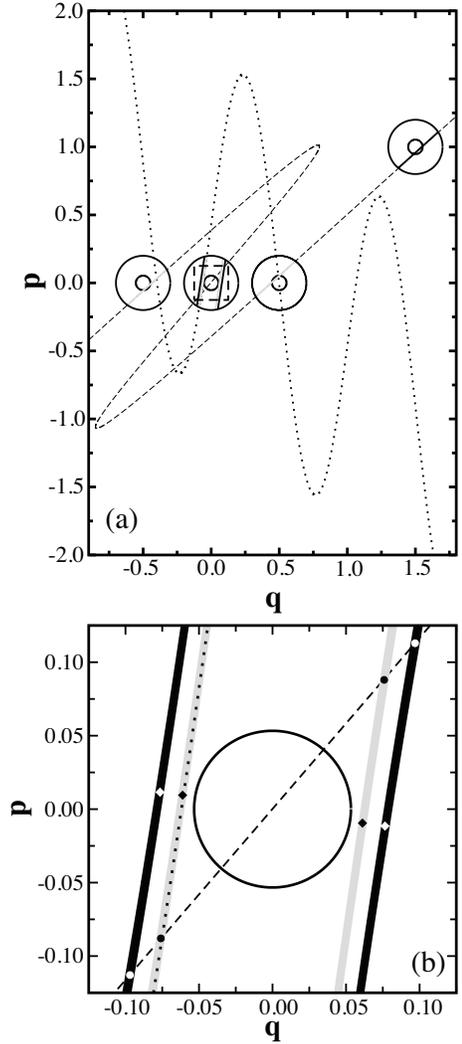}
\caption{ Phase space structure of a chaotic example.  In the upper panel, the unstable manifold of the initial Gaussian density at $(p_\alpha, q_\alpha) = (0,0)$ (dashed line) is propagated two iterations and overlapped with the Gaussian density at $(p_\beta, q_\beta) = (0.0,0.5)= (1.0,1.5)= (0.0,-0.5)$ (torus repetitions of the same density).  The steep branches inside the initial density squeeze, stretch, and translate into the shapes and locations seen by the branches in the final density.  The stable manifold of the final density (dotted line) shows the orientation of the branch inside the initial density.  In the lower panel is an expanded view of the initial wave packet's phase space inside the dashed square.  Two heteroclinic trajectory initial conditions denoted by dots inside their respective branches are $(p_0,q_0)_1=(-0.0892369, -0.0766275)$ and $(p_0,q_0)_2=(-0.1125783, -0.0966593)$ (the other two are found by reflection through the origin).  The saddle point trajectory initial conditions are $({\cal P}_0,{\cal Q}_0)_1=(0.0095152-i 0.0611558, -0.0611558-i 0.0095152)$ and $({\cal P}_0,{\cal Q}_0)_2=(0.0115409 -i 0.0764952, -0.0764952-i 0.0115409)$; the real parts are denoted by diamonds and they are seen to reside in the branches as well.  Inspection reveals ${\cal P}_0= i {\cal Q}_0$ as required by Eq.~(\ref{GGWPDcondition}) for a wave packet centered at $(p_\alpha,q_\alpha)=(0,0)$ with $2\sigma^2=\hbar$.}
\label{fig6}
\end{figure}

Consider a strongly chaotic system with $K=8.25$.  There are two very convenient fixed points of the mapping on which to place the initial and final densities, respectively, and the related wave packets as shown in Fig.~\ref{fig6}a.  Due to the phase space being a torus, in principle all the images of the final density should be shown for a complete figure.  Adding 
any positive or negative integer to the $p$ or $q$-value represents the same phase space point.  Only three images of the final density are shown to simplify the figure.  The solid line circles correspond to the $5\sigma$ contours for the largest and smallest value of $h$.   The heteroclinic trajectories lie on the intersections of the unstable manifold of $(p_\alpha, q_\alpha) = (0,0)$ and the stable manifold of $(p_\beta, q_\beta) = (0.0,0.5)$ (and its images); as just mentioned the images at $(p_\beta, q_\beta) = (1.0,1.5)= (0.0,-0.5)$ represent the same density.  For $t=2$, there are $2$ heteroclinic trajectories (actually $4$ trajectories, but due to reflection symmetry they come in pairs, it simply doubles the semiclassical result).  However, a third heteroclinic trajectory not included in the calculations would be found if the final phase space density centered at $(p_\beta, q_\beta) = (1.0,0.5)$ or $(-1.0,-0.5)$ were used.  The branch represented by this heteroclinic trajectory is further from the center of the initial and final densities and exponentially suppressed compared to the other two (the fact that it is left out affects the accuracy of GGWPD only for the largest values of $\hbar$).  As for the near-integrable case, the Newton-Raphson procedure converged to double precision on the saddle point trajectories with four iterations.  For very strongly chaotic systems and longer time dynamics, the procedure leading to Eq.~(\ref{GGWPDcorrection2}) may prove less reliable because the trajectory is too unstable to follow or even maintain the unit determinant of its stability matrix.  Incorporating techniques, such as found in~\cite{Li15}, may help alleviate this problem.

The two heteroclinic trajectories do not appear to be the best representative trajectories for their respective branches due to the way the unstable and stable manifolds intersect.  This is illustrated in Fig.~\ref{fig6}b, where the phase space density corresponding to the initial wave packet is expanded.  The heteroclinic and real parts of the saddle point trajectories are shown.  The heteroclinic trajectories are a greater number of widths from the center than many other trajectories within their respective branches.  Not too surprisingly, an interesting shift occurs in the real parts of the saddle points relative to the heteroclinic trajectories.  The real parts remain inside their respective branches, but end up very close to the regions where the branches would be tangent to the circular width contours, and thus closest to the center of the density.  

Both semiclassical approximations work very well, much like the integrable case.  The absolute magnitude of the errors of both semiclassical 
\begin{figure}[htb]
\includegraphics[width=3.3in]{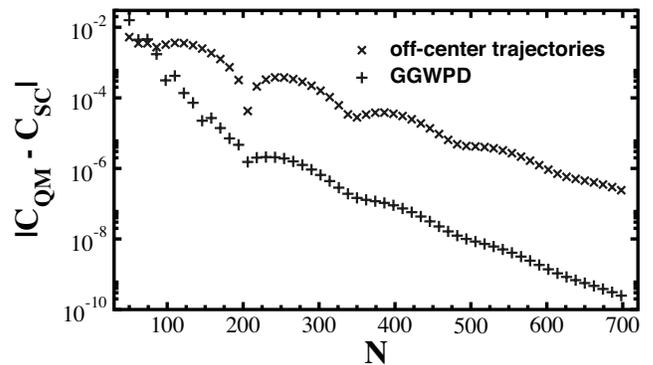}
\caption{Differences of the quantum overlap and semiclassical approximations.  The notation is the same as Fig.~(\ref{fig3}).  The difference magnitudes are shown on a logarithmic scale.  The GGWPD errors are multiple orders of magnitude smaller and an oscillation appears in the errors as the phase relationship between the two contributions varies with $N$.}
\label{fig7}
\end{figure}
methods are illustrated in Fig.~\ref{fig7}.  Again, the errors in magnitude of the difference from the quantum value decrease with shrinking $\hbar$ and the GGWPD errors shrink more rapidly.  The new feature is the introduction of a small oscillation.  As there are two saddle points or two off-center, real trajectories, varying $\hbar$ varies the phase relationship between the two contributions smoothly.  For certain values, the two contributions and their errors destructively interfere and at other values, they constructively interfere.  As $\hbar$ shrinks, one of the contributions becomes more prominent and the oscillations fade.

\begin{figure}[htb]
\includegraphics[width=3.45in]{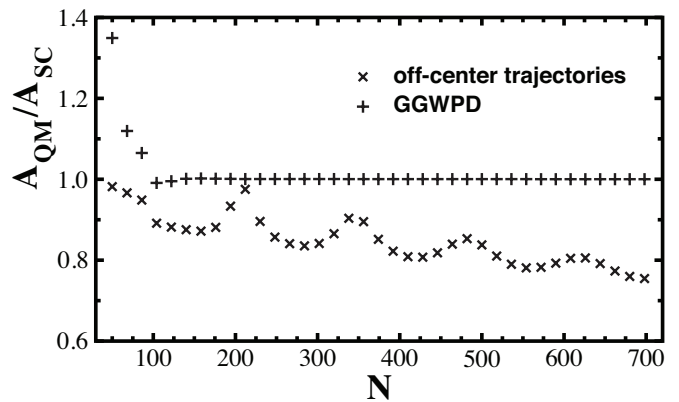}
\caption{ Relative magnitude errors of the two semiclassical methods.  The notation is the same as Fig.~(\ref{fig4}).  The ratio of the quantum magnitude to the semiclassical magnitudes are shown as a function of $N$.  Note that whereas the GGWPD method approaches unity with increasing $N$, the off-center, real trajectory method apparently does not.}
\label{fig8}
\end{figure}
Consider the relative magnitude errors shown in Fig.~\ref{fig8}.  The heteroclinic trajectory sum has a magnitude that diverges from the quantum result, unlike the GGWPD result, even though the absolute errors are shrinking.  This is similar to the integrable example, except a bit worse since there the integrable example appeared to approach a constant relative error, just not converge.  
\begin{figure}[htb]
\includegraphics[width=3.45in]{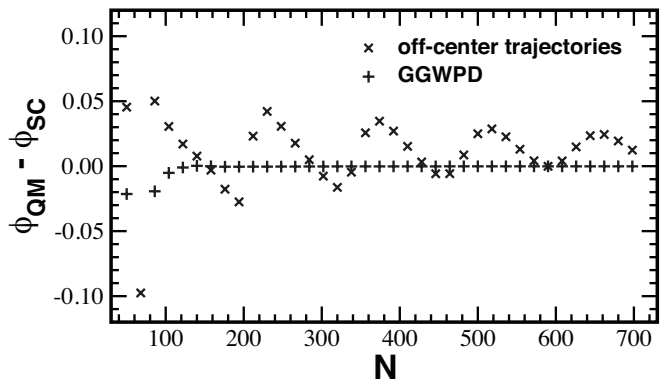}
\caption{ Phase errors of the two semiclassical methods.  The notation is the same as in Fig.~(\ref{fig5}).  The GGWPD phase error vanishes quickly in the semiclassical limit, but the heteroclinic trajectories' errors go through interference oscillations of decreasing magnitude with $\hbar$.  Unlike the integrable example, there does not seem to be a small phase drift.}
\label{fig9}
\end{figure}
Finally, consider the phase errors shown in Fig.~\ref{fig9}.  The interference oscillations of the two contributions with $\hbar$ are most prominent in this case for the off-center, real trajectory method.  In fact, is not entirely clear if the heteroclinic trajectory sum phase error is vanishing, but it is quite small.  On the other hand, the GGWPD result rapidly converges to the correct phase without the phase error oscillations seen for the off-center, real trajectory method.

\section{Conclusions and open problems}
\label{conc}

Without adding corrections for diffraction and uniformizations to account for coalescing saddle points, the ultimate semiclassical theory for propagating wave packets is GGWPD.  It entails carrying out in full details, the method of steepest descents applied to such quantities as $\phi (\vec x;t)$ and $\langle \beta | U_H(t) | \alpha \rangle$.  The latter quantity, in some circumstances, can be considered the same as part of a coherent state representation of the Feynman path integral~\cite{Klauder78} and has been the subject of many studies; see~\cite{Novaes05} and references therein.  The method has the potential to give very accurate results for a wide variety of physical problems that are effectively in a short wavelength limit, and in addition, the potential to give very physical interpretations of the essential physics in such problems.

The GGWPD method has been in existence for more than 25 years~\cite{Huber88}, but has not developed into a widely used and practical technique.  The barriers to its direct implementation are considerable.  Most important are the difficulties of analytically continuing classical dynamics in a complex domain where both position and momenta are complex.  It requires high dimensional root searches in multiple degrees of freedom with quantities exhibiting highly complicated functional behaviors.  It also requires determining whether a particular saddle should be kept or dropped.  However, by recognizing that real classical dynamics imprints itself on the complex dynamics, it is possible to develop indirect root search methods that are vastly easier to implement and avoid several of the pitfalls that a direct root search method would have to confront.  The research contained in~\cite{Vanvoorhis02, Vanvoorhis03, Deaguiar05} are noteworthy in this regard for their advances and understanding, and they rely explicitly or implicitly on the notion of a complex trajectory being ``nearly real.'' 

Although the concept of nearly real may be important, it is also rather nebulous.  We adopt a different perspective, but for some trajectories it is clearly related, that of real crossing saddles.   The representative, off-center, real trajectory methods introduced more than twenty years ago~\cite{Tomsovic91b, Oconnor92,Tomsovic93,Barnes93,Barnes94} sidestepped the complex trajectories issue raised by GGWPD~\cite{Huber88}.  It is worth emphasizing that the off-center, real trajectory methods have no connection to initial value representations~\cite{Heller81, Herman84, Heller91b, Kay94}.  They begin by giving complete classical transport solutions of initially localized phase space densities in integrable and chaotic systems.  

With this complete set of representative trajectories, each relating to a distinct classical transport process, carrying out wave packet dynamics leads to a Gaussian integral for each representative trajectory that can be considered the construction of an approximate saddle, but without going through the trouble of locating the actual saddle point.  The saddle points of those approximate saddles are complex, but there exists a line of real trajectories in each integral that crosses the saddle, and the notion of complex trajectories was entirely avoided.  Here, we have shown that there is a one-to-one correspondence between the approximate representative trajectory saddles and a subset of the true complex saddles of GGWPD, and that these saddles are easily found with a Newton-Raphson scheme if the branches are locally linear.  Since, the representative trajectory method gives a complete transport solution, that accounts for a complete set of saddles associated with classically allowed processes.  Any remaining saddles not found with in this way are necessarily related to classically non-allowed processes, such as tunneling, or are to be rejected for being on the wrong side of Stokes surfaces.  The saddles found with the representative, real trajectories are always on the good side.

One of the most vexing problems of root searching is identifying the set of basins of convergence within which there lies a single root.  These basins may be extremely small or buried deep in a space of enormous volume.  Nevertheless, generally speaking, once a single point within a basin of attraction is identified, using it to initiate a recursive search algorithm tends to converge rapidly to the desired root.  From this perspective, the representative, real trajectories provide a meaningful way to generate one and only one point either within or close to every basin of convergence of a unique saddle, so long as the saddle point is associated with a classically allowed transport process.  If the trajectory is only close to the basin of convergence, an additional initial step must be added to the Newton-Raphson scheme.  This not only greatly simplifies the root search problem, it inextricably links the physical interpretation of the real trajectory to the saddle point trajectory within whose basin of attraction it lies.  It may be rather difficult to find an interpretation of a complex trajectory without this connection.  

The relationship between off-center, real trajectories and saddle points is illustrated here with the kicked rotor for a number of reasons.  First, by changing kicking strength, it is possible to have an integrable (or nearly-integrable) system and a strongly chaotic one.  Second as a map, it provides an extremely simple example that allows for a very detailed view of the how everything works.  For a continuous time system, if one identifies a saddle point at one time, then as time changes the saddle point trajectory deforms continuously and has to be followed as a function of time.  This is straightforward due to continuity (except in presumably rare cases passing through bifurcations or crossing branch cuts), but adds an unnecessary level of difficulty for initially demonstrating the off-center/saddle point  (real-crossing) connection.     Third, it was possible for simplicity to use convenient fixed points and arrange for only one or two off-center, real trajectories to be involved.  Finally, it was possible to compare and contrast the quality of the two approximations as a function of $\hbar$.  Not surprisingly, although both methods worked very well, GGWPD improved the results greatly, even though the cases treated had a local dynamics with very linear branches.  The off-center, real trajectory method contained a couple of novel features in its errors that were cured in the GGWPD approach.  The off-center, real trajectory method's degradation of the GGWPD method would become rather important in a dynamical system where the off-center, real trajectory branches exhibited more curvature.  The greater the curvature of the real trajectory manifolds, the greater the degradation of the GGWPD result.

Within a broad range of semiclassical approximations, it is often taken for granted that, in practice, initial value representations are superior to methods that require root searches.  This is not necessarily so.  It depends on the problem, the goals (e.g.~whether a ``black box'' technique is sought), and it depends on the the level of sophistication that can be applied to the root search.  There are always vastly fewer representative, off-center, real trajectories or saddle points needed than the number of trajectories required for initial value representations.  In fact, it is not surprising to the authors that systems, such as the strongly chaotic stadium billiard where the off-center branches or saddle points explode in number exponentially fast, have only been treated by off-center, real trajectory methods, not initial value representations.  Using the method of this paper continuing with the stadium example, GGWPD would be both straightforward to implement, take almost no additional computational time, and be more accurate than the original heteroclinic summation technique~\cite{Tomsovic93}.  Even the far simpler Coulomb problem displaying revivals and fractional revivals that are treatable semiclassically~\cite{Barnes94} may not have been done yet with an initial value representation method.  In addition, the fact that either a representative or saddle point trajectory represents a branch of trajectories gives a clear physical interpretation.  Each representative (saddle point) trajectory reflects a unique transport pathway.  As seen in the Coulomb revival problem~\cite{Barnes94}, after 20 periods of motion, a transport branch for 17, 18, 19,... 23 collisions with the nucleus all return with a $2\pi$ phase shift between branches of consecutive collision number.    One might say as a shorthand that 7 orbits are sufficient to explain the first revival, but the true meaning is that there are 7 contributing transport pathways (branches) changing continuously with time in the time neighborhood of the revival.  Initial value representation methods do not automatically provide interpretations of this kind, and if one goes through the process of classifying the transport pathways with their trajectories, then one has done all the essential work required for a root search method.

There are a number of further directions suggested by this work.  Wave packet propagation shows up in an incredibly broad range of systems.  Applying the method to interesting physical systems possessing multiple degrees of freedom is an obvious extension to pursue.  That would open up many new possibilities.  Although, we did not give such an application here, the necessary expressions were given.  Possible systems of interest are not limited to quantum ones as other wave mechanical systems, such as found in acoustics and optics, also provide many important applications.  Some wave systems are amenable to paraxial optical approximations in which the results here would apply directly.  Otherwise, new expressions may be derived for the propagation proceeding via the wave equation.  Another direction that would be very interesting is to develop a method to locate tunneling saddle point trajectories without having to use a direct, complex approach.  Despite their not being related to classically allowed processes, there may yet exist methods to locate them with a sufficient understanding of the connections between real and complex classical dynamics.

\appendix
\section{}
\label{append}

The expression for the representative, off-center, real trajectory method needed for the kicked rotor section is given here for completeness.  It is, 
\begin{widetext}
\begin{eqnarray}
\barr{rcl}
\left\langle \beta|U_H(t)| \alpha\right\rangle & = & \displaystyle\sum_{\gamma}\left\langle \beta|U_\gamma(t)| \alpha\right\rangle  =  \sum_{\gamma} \l( \dis\frac{2}{A_0} \r)^{1/2} \exp \l[ \dis\frac{i}{\hbar} \l\{ S(x_t^\gamma,x_0^\gamma)
+ p_t^\gamma \l( x_\beta - x_t^\gamma \r)  - p_0^\gamma (x_\alpha - x_0^\gamma) \r\} - \dis\frac{i\pi\nu}{2} \r.
\\ \\
& - & \l. \dis\frac{1}{2A_0} \l\{ A_1 \delta x_\alpha^2 + A_2 \delta x_\beta^2 + A_3 \delta p_\alpha^2 + A_4 \delta p_\beta^2 - 2 (\delta x_\alpha + i \delta p_\alpha)
(\delta x_\beta - i \delta p_\beta) \r. \r.
\\ \\
& + & \l. \l. 2 i A_1 \delta x_\alpha \delta p_\alpha - 2i A_2 \delta x_\beta \delta p_\beta\r\} \r] \;;
\\ \\
A_0 & = & M_{11} + M_{22} + i \l( \dis\frac{\hbar M_{21}}{2\sigma^2} - \dis\frac{2\sigma^2 M_{12}}{\hbar} \r) \;, \;\;\;\;
A_1 = M_{22} - \dis\frac{2i \sigma^2 M_{12}}{\hbar} \;,
\\ \\
A_2 & = & M_{11} - \dis\frac{2 i  \sigma^2 M_{12}}{\hbar}\;, \;\;\;\; A_3 = M_{11} + \dis\frac{i \hbar M_{21}}{2 \sigma^2} \;,\;\;\;\;
A_4 = M_{22} + \dis\frac{i \hbar M_{21}}{2 \sigma^2}\;,
\\ \\
\delta x_\alpha & = & (x_\alpha - x_0^\gamma)/\sqrt{2\sigma^2}\;,\;\;\;\; \delta p_\alpha = (p_\alpha-p_0^\gamma)\sqrt{2\sigma^2}/\hbar\;, 
\\ \\
\delta x_\beta &  = &  (x_\beta - x_t^\gamma)/\sqrt{2\sigma^2}\;,\;\;\;\; \delta p_\beta = (p_\beta-p_t^\gamma)\sqrt{2\sigma^2}/\hbar\;.
\earr
\label{eq.14}
\end{eqnarray}
\end{widetext}

\acknowledgments

The authors gratefully acknowledge helpful discussions with and a critical reading of the manuscript by Arul Lakshminarayan, Lucas Kocia, and Eric Heller.  Financial support of this work was generously provided by the US National Science Foundation (Grant No. PHY-0855337), CONACyT Project No. 154586, and UNAM-DGAPA-PAPIIT Project nos. IN100803 and IG101113.  HP and MV gratefully acknowledge postdoctoral fellowships from DGAPA-UNAM.

\bibliography{general_ref,molecular,quantumchaos,classicalchaos,rmtmodify}

\end{document}